\documentclass[12pt]{article}
\usepackage{graphicx}
\usepackage{amsmath}
\newcommand\pubnumber{SNSN-XXX-YY}
\newcommand\pubdate{\today}

\def\napoli{Department of Physics\\
New Mexico State University, PO Box 30001, MSC 3D, Las Cruces, NM 88003, USA}
\def\Title#1{\begin{center} {\Large #1 } \end{center}}
\def\Author#1{\begin{center}{ \sc #1} \end{center}}
\def\Address#1{\begin{center}{ \it #1} \end{center}}

\newcommand\pubblock{\rightline{\begin{tabular}{l} \pubnumber\\
         \pubdate  \end{tabular}}}
\newenvironment{Abstract}{\begin{quotation}  }{\end{quotation}}
\newenvironment{Presented}{\begin{quotation} \begin{center} 
             PRESENTED AT\end{center}\bigskip 
      \begin{center}\begin{large}}{\end{large}\end{center} \end{quotation}}

\begin{document}
\begin{titlepage}
\pubblock

\vfill
\Title{Exploring Neutrino Interactions with MicroBooNE}
\vfill
\Author{ Tia Miceli for the MicroBooNE Collaboration}
\Address{\napoli}
\vfill
\begin{Abstract}
Recently, MiniBooNE observed an electromagnetic excess at low energy.
What is the nature of this excess? What about the nature of the low-energy excess at LSND 20 years ago?
The MicroBooNE detector will see neutrinos from the same Booster beam at Fermilab as used by MiniBooNE.
MicroBooNE's design will enable us to discriminate photons from electrons elucidating the MiniBooNE and LSND low-energy electromagnetic excesses.
MicroBooNE is a 170 ton liquid argon time projection chamber (LArTPC) capable of imaging neutrino interactions with the detail of a bubble chamber, but with electronic data acquisition and processing.
In addition to shining light on the low-energy excesses and measuring low-energy neutrino cross sections, MicroBooNE is leading the way for a more extensive short-baseline neutrino physics program at Fermilab and it also serves as a R\&D project towards a long-baseline multi-kiloton scale LArTPC detector.
\end{Abstract}
\vfill
\begin{Presented}
XXXIV Physics in Collision Symposium \\
Bloomington, Indiana,  September 16--20, 2014
\end{Presented}
\vfill
\end{titlepage}
\def\thefootnote{\fnsymbol{footnote}}
\setcounter{footnote}{0}
%

%%%%%%%%%%%%%%%%%%%%%%%%
%  Introduction                                                 %
%%%%%%%%%%%%%%%%%%%%%%%%
\section{Introduction}

MicroBooNE is a new detector capable of recording neutrino interactions in great detail and quickly digitizing the results for analysis.
This is the sort of detector that is needed to reduce backgrounds that limit other experiments.
MicroBooNE's design enables excellent electron and photon detection allowing studies of the MiniBooNE low-energy excess in the reconstructed electron flavor neutrino and antineutrino spectra.
With experience being gained in this new technology, further deployments are planned to explore  short- and long-baseline neutrino oscillations.

Here we summarize the motivation and general detection principles of this new technology.
We also show possible outcomes of probing the short-baseline low-energy electron flavor neutrino spectrum.

%%%%%%%%%%%%%%%%%%%%%%%%
%  Motivation                                                    %
%%%%%%%%%%%%%%%%%%%%%%%%
\section{Motivation}

\begin{figure}[htb]
\centering
\includegraphics[height=3.2in]{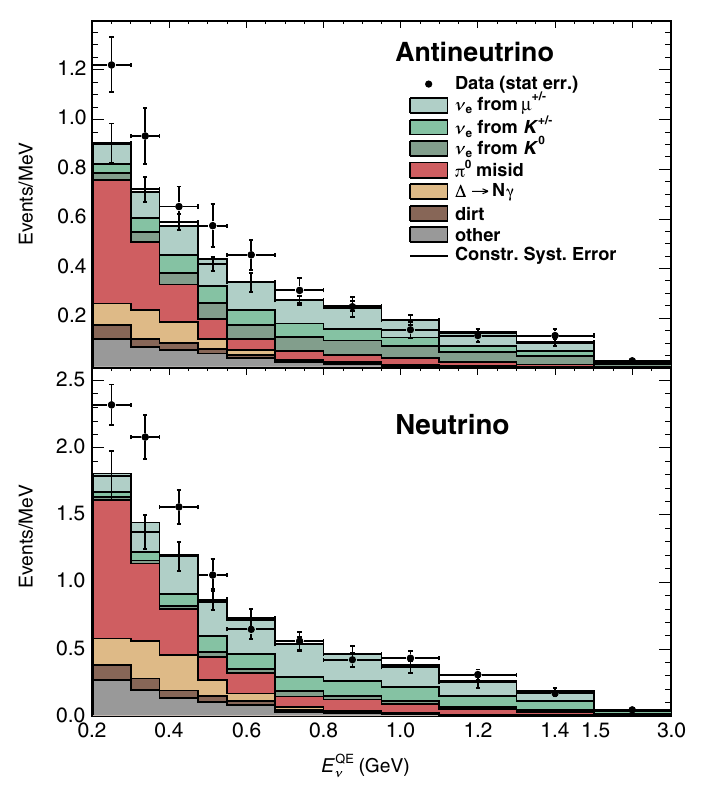}
\caption{The reconstructed quasi-elastic neutrino energy, $E^{ \text{QE} }_{\nu}$ spectra at MiniBooNE for the BNB operating in antineutrino mode (top), and neutrino mode (bottom). Data are shown in black points, unconstrained backgrounds shown in stacked filled histograms, the background systematic uncertainty for constrained backgrounds is shown by the thinner uncertainty bars~\cite{Aguilar-Arevalo:2013pmq}.}
\label{fig:minibooneExcess}
\end{figure}

An earlier experiment, MiniBooNE, measured the electron neutrino energy spectrum in the Booster Neutrino Beamline (BNB) at Fermilab.
The BNB provides primarily muon flavor neutrinos and antineutrinos depending on the selected beam configuration. 
Electron flavor neutrinos in this beam are found by looking for charged current quasi-elastic (CCQE) interactions.
These electron flavor neutrinos arise from oscillations, from beam-intrinsic backgrounds due to muon, kaon, and pion decay, or due to the misidentification of photons as electrons.
The MiniBooNE experiment measured an excess to the expected number of electron neutrinos with low-energy~\cite{Aguilar-Arevalo:2013pmq}, Fig.~\ref{fig:minibooneExcess}.

The neutrino energy reconstructed in CCQE events depends on the energy and scattering angle of the outgoing electron.
In oil Cherenkov detectors like MiniBooNE, the Cherenkov ring detected for an electron is similar to the signal for a photon because it can convert to a collinear electron-positron pair.
Because of the limits of Cherenkov technology, MiniBooNE was unable to discover the nature of the excess in this poorly modeled low-energy region.
A new detector technology is needed to discover the identity of the electromagnetic signal contributing to the excess.
MicroBooNE has excellent photon/electron identification abilities to measure the BNB.

MicroBooNE is the first large liquid argon time projection chamber (LArTPC) in the US.
In addition to resolving the nature of the MiniBooNE excess, MicroBooNE will be able to measure short-baseline neutrino oscillations and the cross section of neutrinos on argon.
The technical experience gained from such liquid argon experiments will inform the design of future experiments.
Several collaborations have come together to propose an extensive multi-detector LArTPC short-baseline neutrino oscillation program at Fermilab, as well as an international long-baseline program.

%%%%%%%%%%%%%%%%%%%%%%%%
%  Experimental Design                                   %
%%%%%%%%%%%%%%%%%%%%%%%%
\section{Experimental Design}

\begin{figure}[htb]
\centering
\includegraphics[height=1.5in]{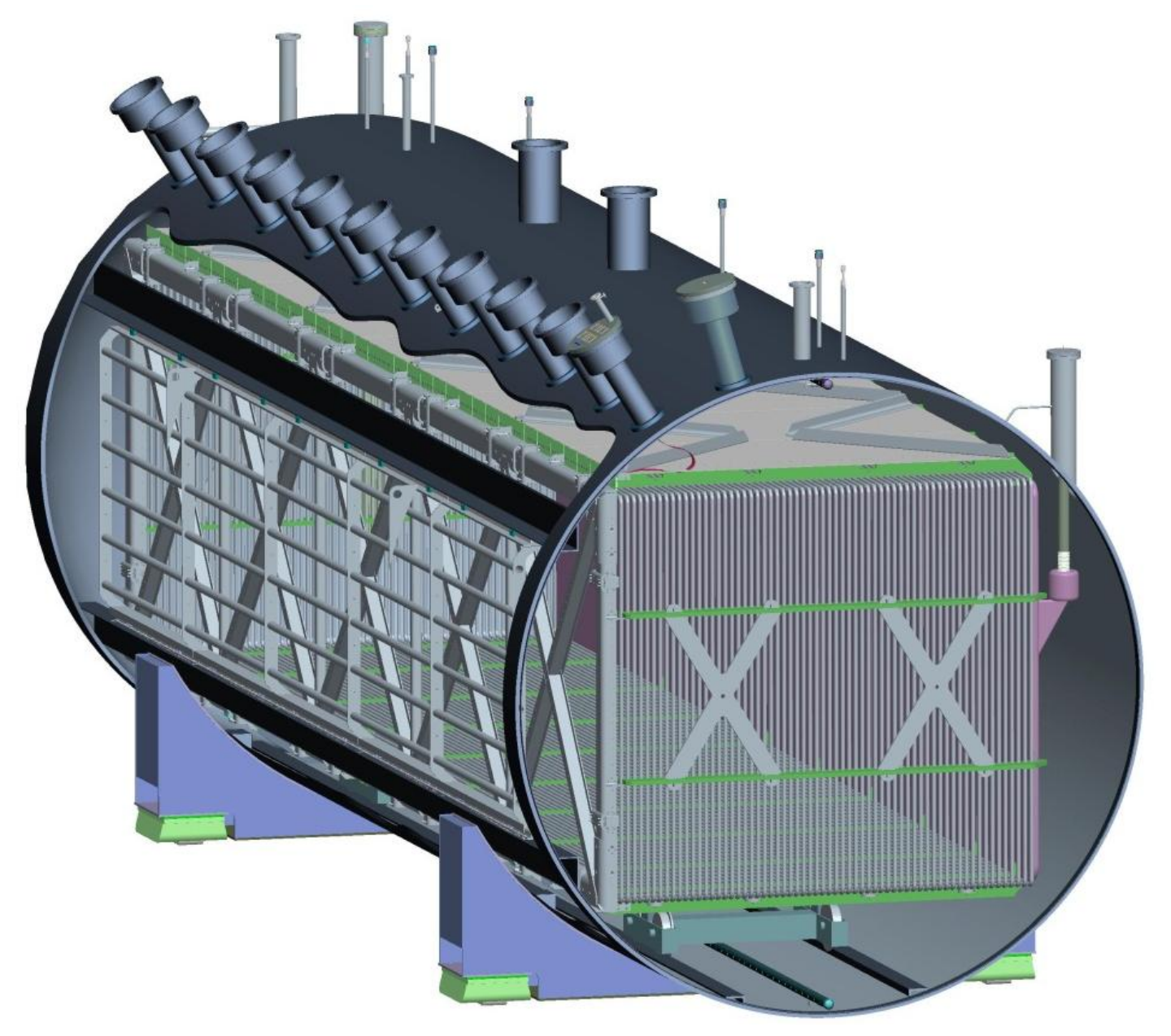} \hspace{15mm} \includegraphics[height=1.5in]{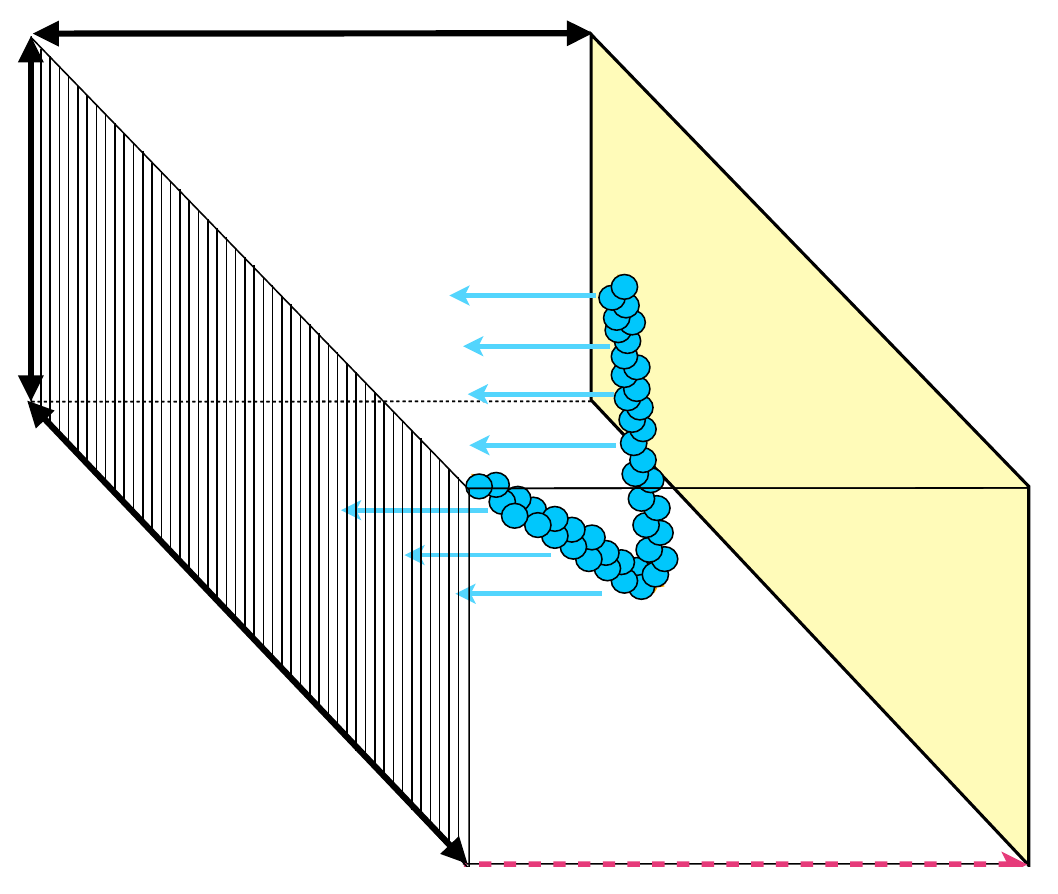}
\caption{Left, a cutaway of the MicroBooNE detector. The cylindrical cryostat holds the liquid argon and the drift chamber.  Right, ionization electrons produced by passing charged particles drift to the left, towards three wire signal planes.}
\label{fig:uB}
\end{figure}

LArTPCs provide uniquely distinguishable signals for photons and electrons, so they are ideal for resolving the ambiguous electromagnetic excess in MiniBooNE. 
MicroBooNE is a 170 ton (total volume) LArTPC located on Fermilab's BNB with a drift distance of 2.5 m, and is 10.5 m long and 2.3 m tall~\cite{tdr}, Fig.~\ref{fig:uB}.
Since MicroBooNE was built to investigate the excess in MiniBooNE, it is important to control variables that may change the experimental conditions.
We want the BNB to retain its same configuration that it had during MiniBooNE data-taking.
The MicroBooNE detector is also situated so that it has the same $L/E$ (1 m/MeV) as the MiniBooNE detector.
As much as possible, we want to replicate the conditions in MiniBooNE so we can uncover their observed excess.

\begin{figure}[htb]
\centering
\includegraphics[height=1.2in]{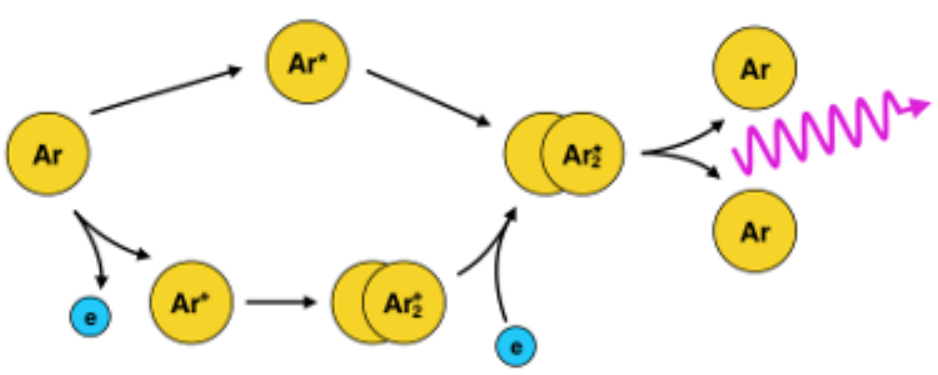}
\caption{Argon atoms become excited (top) or ionized (bottom) when a charged particle passes nearby. An excited argon atom  briefly forms an Ar$_2$ molecule, then the atom de-excites and emits scintillation light. The ionized argon atom, on the other hand, loses an electron that may be detected by the wire plane. That argon ion then may bond with another argon atom, absorb an electron, form an excited Ar$_2$ molecule, then also emit scintillation light as it relaxes. }
\label{fig:Ar}
\end{figure}

\begin{figure}[htb]
\centering
\includegraphics[height=1.5in]{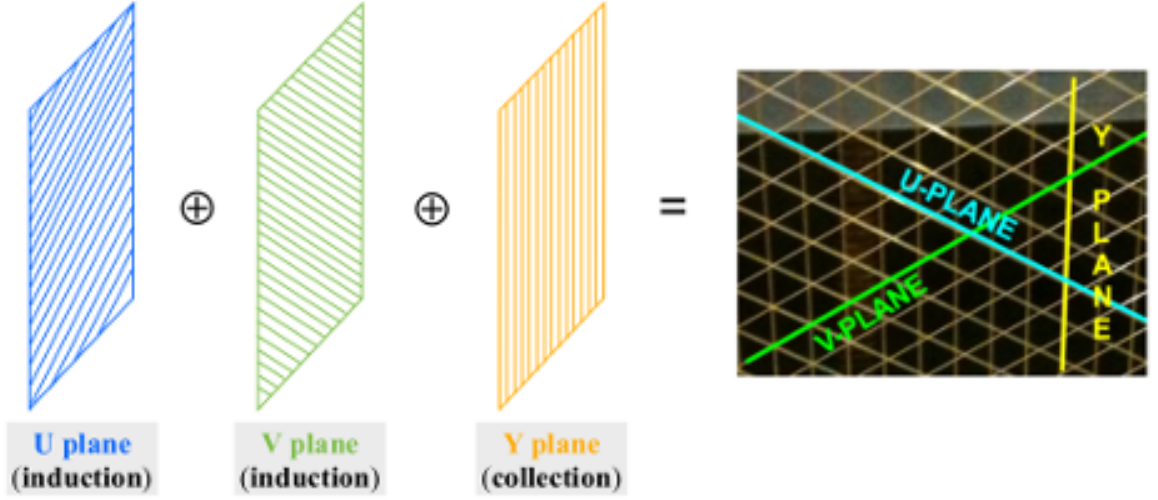} \hspace{10mm} \includegraphics[height=1.5in]{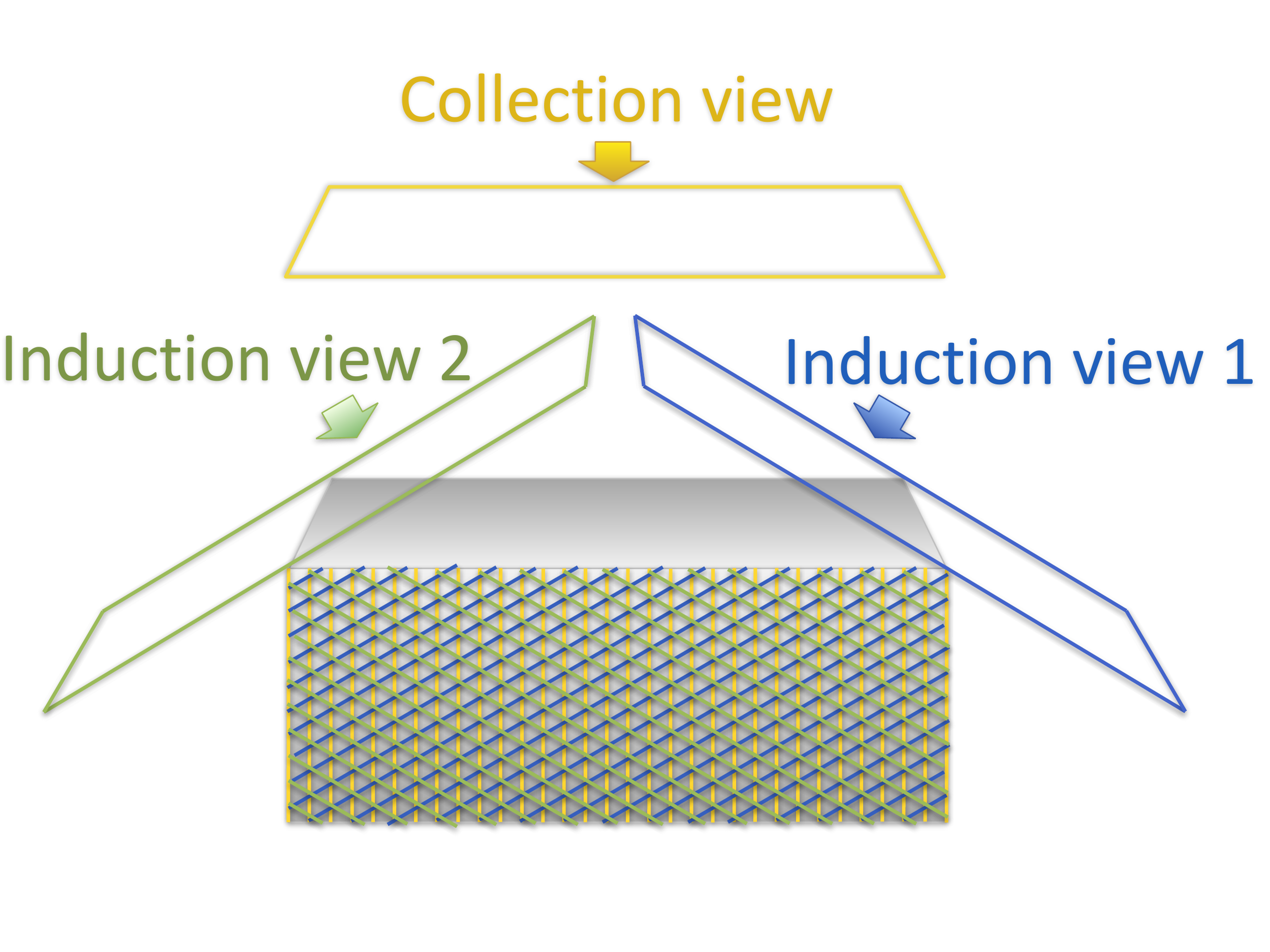}
\caption{Left, zoom in of wire plane. The U and V wires carry an induction signal as the electrons pass on their way to the vertical collection plane, Y. Graphic courtesy Jonathan Asaadi. Right, a visualization of the coordinate position reconstruction in the TPC. Graphic courtesy Georgia Karagiorgi. }
\label{fig:wires}
\end{figure}

As a charged particle passes through the liquid argon in MicroBooNE, a nearby argon atom experiences one of two processes, it becomes ionized, or excited, Fig.~\ref{fig:Ar}.
Most of the ionized electrons will freely travel through the liquid argon in a $\sim500$~V/cm electric field to the TPC anode wires~\cite{tdr}.
The TPC has two induction wire planes at $\pm60^{\circ}$, and one vertical collection plane, Fig.~\ref{fig:wires} left.
The three planes enable us to reconstruct events projected onto a plane parallel with the wires.
The third dimension, along the drift direction, can be resolved by transforming the drift time to the particular depth in the TPC, by using the timing of the beam pulses or photomultiplier tube (PMT) signals~\cite{tdr}.

As the argon atoms are ionized or excited by passing charged particles, both processes may terminate with the argon atom de-exciting via scintillation.
The energy emitted as light can be collected to more carefully identify the passing charged particle.
As argon scintillates, it emits light at 128 nm which cannot be directly detected by the PMTs.
The wavelength of this light must be shifted by a plate mounted in front of each PMT.
The plate is acrylic with a coating of tetra phenyl butadiene to shift the 128 nm scintillation light to 425-450 nm.
This new wavelength is detected by one of the 32 eight-inch PMTs.
Measuring the amount of scintillation in addition to the amount of ionization will improve the $dE/dx$ measurements and improve particle identification.

In such a time projection chamber, photons and electrons, as well as other particle types, may be easily resolved using $dE/dx$.
For example, an electron will have a track with a $dE/dx$ of a minimum ionizing particle (MIP) until it begins to produce an electromagnetic shower.
A high energy photon, however, will not leave a trace until it pair converts or Compton scatters, leaving a displaced vertex.
The track after the photon interacts may have the $dE/dx$ of a MIP, in the case of Compton scattering, or a $dE/dx$ of two MIPs in the case of conversion to an electron and positron.
The excellent particle identification capabilities of LArTPCs make them ideal to resolve the nature of the MiniBooNE excess.

%%%%%%%%%%%%%%%%%%%%%%%%
%  Analysis                                                       %
%%%%%%%%%%%%%%%%%%%%%%%%
\section{Analysis}

The three wire planes of the TPC collect ionization of tracks and showers in the argon enabling a 2D reconstruction on the plane parallel with the wires.
The third dimension is determined by transforming drift time, into a drift distance, Fig.~\ref{fig:wires} right.
%The reconstruction programs do the same job that our brain does when we look at a 3D anaglyph, Fig.~\ref{fig:3d}.

%\begin{figure}[htb]
%\centering
%\includegraphics[height=3in]{views}
%\caption{ Visualization of the coordinate position reconstruction in the TPC. Graphic courtesy Georgia Karagiorgi. }
%\label{fig:views}
%\end{figure}

%\begin{figure}[htb]
%\centering
%\includegraphics[height=2in]{3d}
%\caption{ A cartoon 3D anaglyph of a Neutral Current event in the MicroBooNE drift volume. A neutrino coming in from the right and exchanges a Z boson with the quarks in a proton. The collision results in a recoiling nucleus, short stub in the upper right, a medium length track on the bottom right, likely a charged pion, and the long shower on the left, a converted photon, likely from a neutral pion. Special thanks to Alistair McLean, NMSU, for the MicroBooNE computer model screen captures. Event superimposed from ArgoNeuT~\cite{AndreTalk}. }
%\label{fig:3d}
%\end{figure}

Particles are identified in the LArTPC by getting the energy along the trajectory ($dE/dx$) in the form of ionization electrons and scintillated light.
With the particle identification, energy, and kinematical geometry the interaction can be reconstructed.

\begin{figure}[htb]
\centering
\includegraphics[height=1.8in]{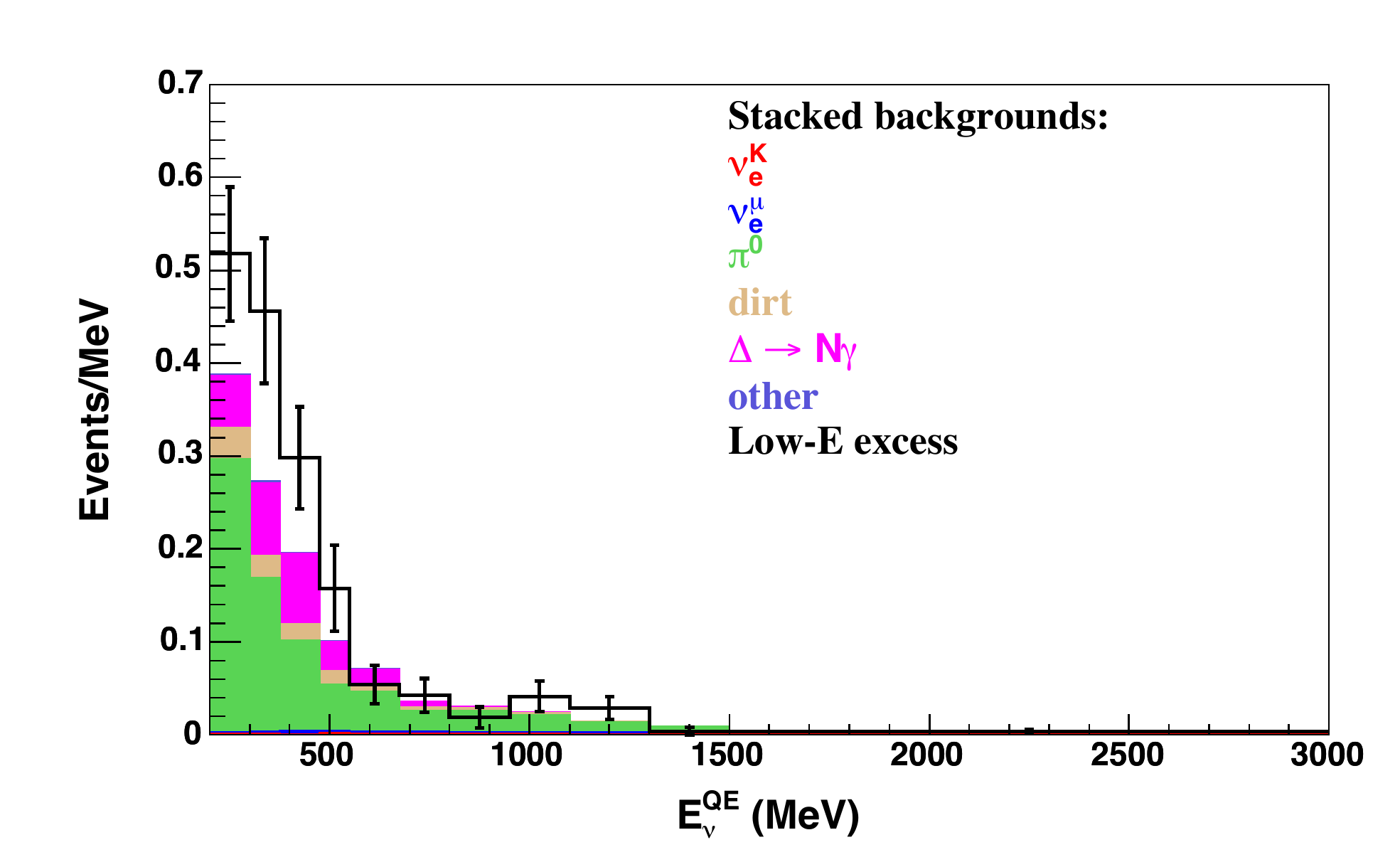} \hspace{2.25mm} \includegraphics[height=1.8in]{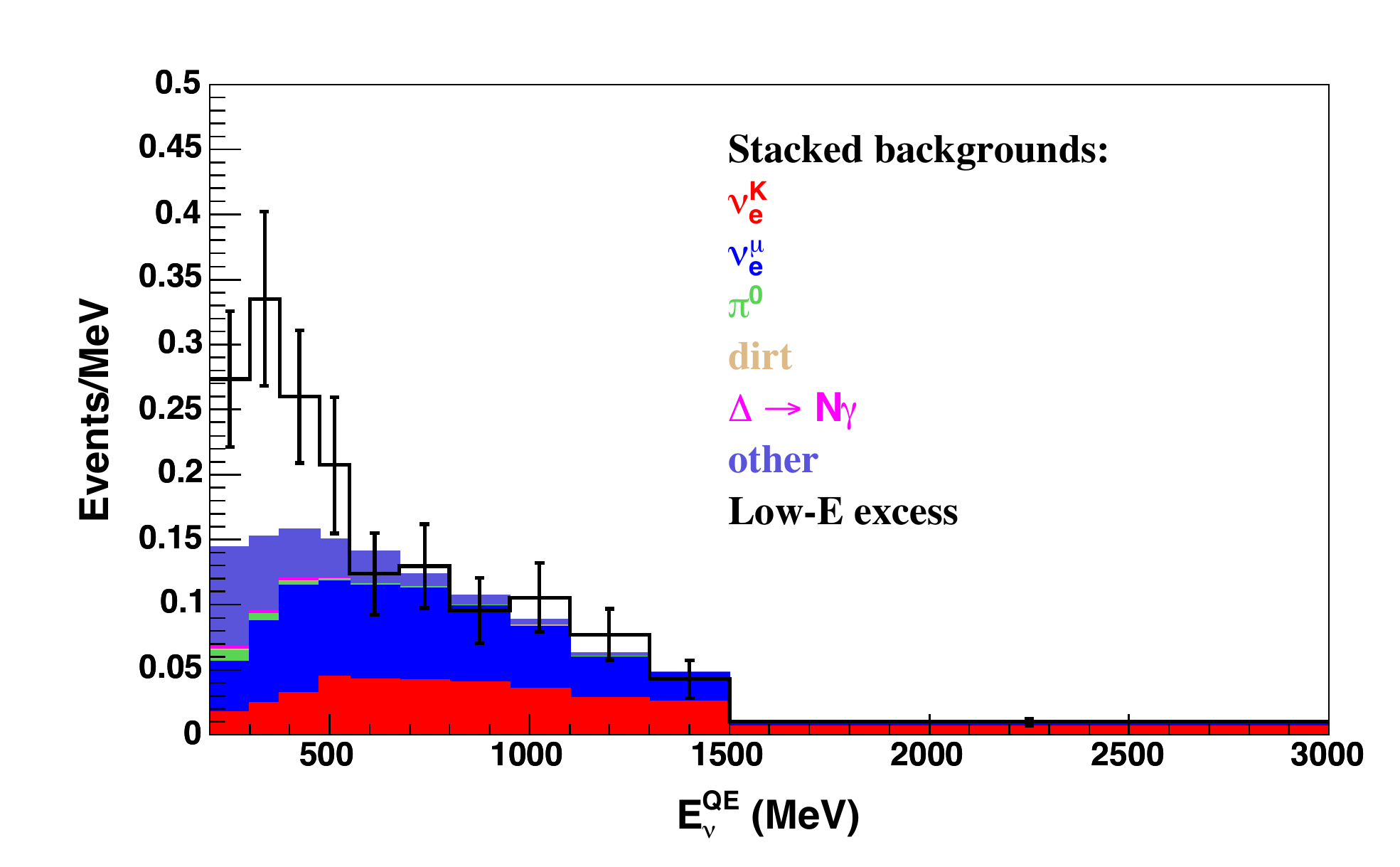}
\caption{Predicted reconstructed energy spectrum for two separate analyses. The analysis optimizing photon signal over background (left) and electron signal over background (right). See description in text for more detail~\cite{tdr}.}
\label{fig:future}
\end{figure}

If the observed MiniBooNE electromagnetic excess is due to photons, an analysis will be optimized to measure photon signal over background to give a spectrum similar to Fig.~\ref{fig:future} left.
The black line assumes the low energy excess observed by the MiniBooNE detector scaled to the MicroBooNE detector.
The event predictions account for assumed MicroBooNE detector efficiency, fiducial volume, and assumed electron-photon separation efficiency.
The prediction assumes data collected for $6.6 \times 10^{20}$ protons on target in neutrino mode, approximately three years of data-taking.
The error bars indicate statistical-only uncertainty.
The most significant backgrounds for this analysis will be due to events that produce real photons like those with $\pi^0$ or a $\Delta$ resonances decaying to a nucleon and a photon. 

If the electromagnetic excess is instead due to electrons, an analysis will be optimized to measure electrons over background and the most significant backgrounds will be different, as shown in Fig.~\ref{fig:future} right~\cite{tdr}.
Backgrounds which have real electrons will be the most significant contaminant of the data sample due to $\mu$, $K$, $\pi$ decays.
 
%%%%%%%%%%%%%%%%%%%%%%%%
%  Conclusion                                                  %
%%%%%%%%%%%%%%%%%%%%%%%%
\section{Conclusion}

If the excess is dominated by photon events, we will have unveiled a new photon background never before considered in neutrino experiments. 
On the other hand, if the excess is dominated by electron events, we may discover new oscillation modes into a fourth generation neutrino.
Either way, MicroBooNE will probe something new about nature while providing an arena for physicists to gain technical experience as we prepare for the short- and long-baseline programs being planned at Fermilab.

\bibliographystyle{amsplain}
\bibliography{ProceedingsPIC2014_Miceli.bib}
%\bib{MicroBooNETDR}{MicroBooNE Document 1821-v13. The MicroBooNE Collaboration. ÒThe MicroBooNE Technical Design ReportÓ, http://microboone-docdb.fnal.gov/cgi-bin/ShowDocument?docid=1821}
 
\end{document}